\documentclass[letter,longauth]{aa}
\usepackage{graphicx}
\usepackage{hyperref}
\usepackage{amsmath}
\usepackage{amstext}
\usepackage{epstopdf}
\usepackage{setspace}
\usepackage[style=base]{caption}
\usepackage{textcomp,gensymb}
\usepackage[varg]{txfonts}
\usepackage{float}
\usepackage{subfig}
\usepackage{natbib,twoopt}

\bibpunct{(}{)}{;}{a}{}{,}
\newcommandtwoopt{\citeyearads}[3][][]%
{\href{http://adsabs.harvard.edu/abs/#3}{\citeyear[#1][#2]{#3}}}
\hypersetup{colorlinks,linkcolor=red,citecolor=blue,urlcolor=blue}

\newcommand{\alfe}{[$\alpha$/Fe]}
\newcommand{\mh}{[M/H] }
\newcommand{\ratio}{$\alpha_\mathrm{poor}$/$\alpha_\mathrm{rich}$ }
\begin{document}
\title{The Gaia-ESO Survey: $\alpha$-abundances of metal-poor stars} 
\author{R. Jackson-Jones\inst{\ref{inst1}}
\and
P. Jofr\'{e}\inst{\ref{inst1}}
\and
K. Hawkins\inst{\ref{inst1}}
\and
A. Hourihane\inst{\ref{inst1}}
\and
G. Gilmore\inst{\ref{inst1}}
\and
G. Kordopatis\inst{\ref{inst1}}
\and
C. Worley\inst{\ref{inst1}}
\and
S. Randich\inst{\ref{arcreti}}
\and
A. Vallenari\inst{\ref{inst3}}
\and                
T. Bensby\inst{\ref{inst4}}
\and
A. Bragaglia \inst{\ref{bol}}
\and
E. Flaccomio\inst{\ref{inst6}}
\and
A. J. Korn\inst{\ref{inst8}}
\and
A. Recio-Blanco\inst{\ref{inst2}}
\and
R. Smiljanic\inst{\ref{torun}}
\and
M. T. Costado\inst{\ref{inst7}}
\and
U. Heiter\inst{\ref{inst8}}
\and
V. Hill\inst{\ref{inst2}}
\and
C. Lardo \inst{\ref{liverpool}}
\and
P. de Laverny\inst{\ref{inst2}}
\and
G. Guiglion\inst{\ref{inst2}}
\and
S. Mikolaitis\inst{\ref{inst2},\ref{inst5}}
\and
S. Zaggia \inst{\ref{inst3}}
\and
G. Tautvai\v{s}ien\.{e}\inst{\ref{inst5}}
}
\institute{Institute of Astronomy, University of Cambridge,
Madingley Road, Cambridge CB3 0HA, United Kingdom \label{inst1} 
\and
INAF - Osservatorio Astrofisico di Arcetri, Largo E. Fermi 5, 50125, Florence, Italy \label{arcreti}
\and
INAF - Padova Observatory, Vicolo dell'Osservatorio 5, 35122 Padova, Italy\label{inst3} 
\and
Lund Observatory, Department of Astronomy and Theoretical Physics, Box 43, SE-221 00 Lund, Sweden\label{inst4} 
\and
INAF - Osservatorio Astronomico di Bologna, Via Ranzani 1, I-40127 Bologna, Italy\label{bol}
\and
INAF - Osservatorio Astronomico di Palermo, Piazza del Parlamento 1, 90134, Palermo, Italy\label{inst6} 
\and
Department of Physics and Astronomy, Uppsala University, Box 515, 75120 Uppsala, Sweden\label{inst8}
\and
Laboratoire Lagrange (UMR7293), Universit\'e de Nice Sophia Antipolis, CNRS,Observatoire de la C\^ote d'Azur, CS 34229,F-06304 Nice cedex 4, France\label{inst2}
\and
Department for Astrophysics, Nicolaus Copernicus Astronomical Center, ul. Rabia\'{n}ska 8, 87-100 Toru\'{n}, Poland\label{torun}
\and
 Instituto de Astrof\'{i}sica de Andaluc\'{i}a-CSIC, Apdo. 3004, 18080 Granada, Spain\label{inst7} 
 \and
Astrophysics Research Institute, Liverpool John Moores University, 146 Brownlow Hill, Liverpool L3 5RF, United Kingdom\label{liverpool}
\and
Institute of Theoretical Physics and Astronomy, Vilnius University, A. Gostauto 12, LT-01108 Vilnius, Lithuania\label{inst5} 
}

\offprints{ \\ 
R. Jackson-Jones, \email{rj320@ast.cam.ac.uk},\\ 
P. Jofr\'e, \email{pjofre@ast.cam.ac.uk}}

\date{Received ; accepted }
\abstract
{We performed a detailed study of the ratio of low-$\alpha$ to high-$\alpha$ stars in the Galactic halo as observed by the Gaia-ESO Survey. Using  a sample of 381 metal-poor stars from the second  internal data release, we found that the value of this ratio did not show evidence of systematic trends as a function of metallicity, surface gravity, Galactic latitude, Galactic longitude, height above the Galactic plane, and Galactocentric radius. We conclude that the \ratio\ value of $0.28 \pm 0.08$ suggests that in the inner halo, the larger portion of stars were formed in a high star formation rate environment, and about 15\% of the metal-poor stars originated from much lower star formation rate environments.}

\maketitle

\section{Introduction}

The halo of the Milky Way is thought to contain some of the oldest stars in the Galaxy \citep[e.g.][]{S6,N12b,jofre,  Keith} and as such, the halo is the perfect place to look for clues regarding the formation of the Galaxy.

Type II supernovae produce $\alpha$-elements and Fe on timescales of approximately 20 Myr in regions with high star formation rates, affecting the \alfe\footnote{The $\alpha$-elements are those that are formed from a combination of helium nuclei -- alpha particles  -- including O, Ne, Mg, Si, S, Ar, Ca, and Ti. The notation \alfe\ is commonly used for the averaged abundance of different $\alpha$-elements with respect to iron.} 
value of their environment \citep{wyse}. On the other hand, Type Ia supernovae produce primarily iron-peak elements on timescales that are usually a few Gyr after the formation of the first stars in environments with lower star formation rates. This timescale can be shorter depending on the details of the star formation environment \citep[e.g.][]{Matteucci}. The deposition of this extra iron leads to a reduction in the \alfe\ value of the environment over time. Thus, stellar [$\alpha$/Fe] values can reveal details about early Galactic star formation sites \citep{Freeman}.

The idea of grouping the metal-poor stars by  their elemental abundances, such as the $\alpha-$elements, in order to trace the origin of the halo has been assessed for decades \citep[see][and references therein for a review]{2003A&A...404..187G, 2003A&A...406..131G}. Recent studies on the \alfe\ values of stars in the Milky Way halo have been reported by, among others, \cite{N10,N11,N12a}, \cite{N12b}, \cite{Sheffield}, \cite{Ramirez} and \cite{Keith}. Similar studies of the \alfe\ value of large samples of disk stars have also been performed by \cite{Lee, Schlesinger, Cheng}. In particular, by studying nearby stars (d $\lesssim$ 330 pc), \cite{N10} concluded that two halo populations exist, an $\alpha_\mathrm{poor}$ (\alfe $<$ +0.3) and an  $\alpha_\mathrm{rich}$ (\alfe $>$ +0.3) population which are distinct in age and kinematics. However, the fact that all these stars are nearby means they may not be representative of the entirety of the halo. The $\alpha_\mathrm{poor}$ stars are interpreted as accreted stars, and the $\alpha_\mathrm{rich}$ stars part of an \emph{in situ} population.  Typically, low metallicity stars are expected to be $\alpha_\mathrm{rich}$. In the \alfe-[Fe/H] diagram, the \alfe\ value is constant and starts decreasing as a function of increasing metallicity. This happens at an [Fe/H] value which depends on the mass of the star forming gas cloud so it is usually found at higher metallicities for dwarf spheroidals compared to the Milky Way \citep[e.g.][]{1997ARA&A..35..503M}.  

A ratio of the number of stars in these two putative populations (\ratio) allows us to investigate the importance of accreted material in the formation of the Milky Way. In this study, we use the metal-poor stars of the second internal data release of the Gaia-ESO Survey \cite[GESviDR2, ][hereafter, GES]{GES, randich} in order to investigate the \ratio ratio for the halo. The selection criteria for field stars of GES consist only of a cut in colour-magnitude space, resulting in a sample that is unbiased in \alfe\ \citep[see][for the selection criteria]{GES}. \cite {2014A&A...567A...5R} and \cite{iDR1} recently noted that in the first internal data release, 10\% of the metal-poor stars are $\alpha_\mathrm{poor}$.
Our purpose is to extend this work by exploring the \ratio value as a function of stellar metallicity, $\log g$, Galactic longitude, Galactic latitude, height above the Galactic plane, and distance from the Galactic centre, as well as by using more data from GESviDR2.

\section{Data}
\subsection{The Gaia-ESO survey}
The GES is an ongoing spectroscopic survey that will observe more than $10^5$ stars, using the FLAMES/GIRAFFE instrument at the ESO.
This work uses two grating set-ups -- HR10 ($R\sim19800$) and HR21 ($R\sim16200$) -- of the GIRAFFE spectrograph with wavelength ranges 5339\AA\ - 5619\AA\ and 8484\AA\ - 9001\AA, respectively.  The spectra are reduced as described in Lewis et al. (in prep). 
Stellar parameters are derived from the spectra by the GES consortium via a combination of three methods: MATISSE \citep{2006MNRAS.370..141R}, FERRE \citep[ and further developments]{2006ApJ...636..804A}, and SME \citep[ and further developments]{1996A&AS..118..595V}.  The parameters are calibrated to perform well with a set of pre-defined clusters (Pancino et al. in prep.) as well as with the {\it Gaia} benchmark stars \citep[Heiter et al. in prep.; ][]{2014A&A...564A.133J}.  For a detailed description of this analysis, see Recio-Blanco et al. (in prep.).  
To select halo stars we considered only the stars with a global metallicity \mh $< -1$  which was determined from spectra with a signal-to-noise ratio (S/N) larger than 10.  We expect to have a small ($\sim 3$\% of the entire thick disk) contamination of metal-poor thick-disk stars in this sample \citep{Kordo2013}.
Finally, stars with recommended \alfe~$> +0.6$  were rejected as they typically had very large errors in the parameters. Although we are aware that we might have missed real very $\alpha_{rich}$ stars, these cases were very few and did not affect our final results. For our aim of studying  the \ratio\ value, we preferred to be conservative and remove all stars with \alfe~$> +0.6$. Our final sample contained 381 stars, with a mean S/N of 22, and median errors of 77~K, 0.19~dex, 0.1~dex and 0.07~dex in effective temperature, surface gravity, metallicity, and \alfe, respectively.  
\subsection{Confirming halo membership}
Although GES has proper motion data for several targets, the stars in this study are fainter than the bulk of the GES data so most of their proper motions are very uncertain. In the absence of full kinematics, as a further halo membership check we investigated the infrared colour-magnitude diagram of the sample. 
\begin{figure}[t]
\includegraphics[width=1.15\linewidth]{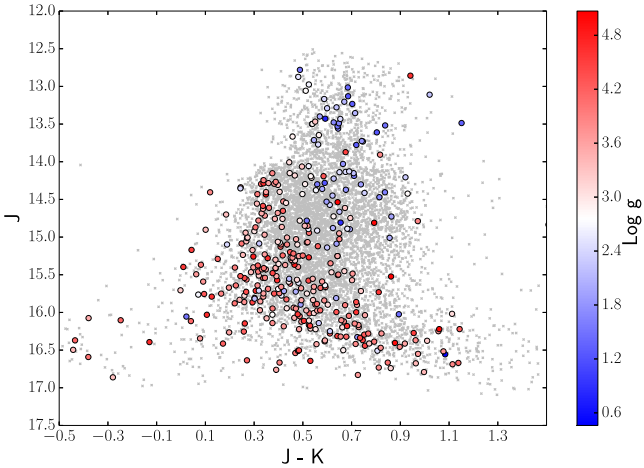}
\caption{\tiny The extinction-corrected 2MASS $J$ magnitudes and 2MASS $J-K$ colours for the 381 stars of our sample coloured as a function of $\log g$. The rest of the metal-rich stars from the GESviDR2 are coloured in grey.}
\label{fig:HR}
\end{figure}
We corrected our $J$ and $K$ magnitudes using the extinction corrections of \cite{Yuan}. In Fig.~\ref{fig:HR} we display the dereddened magnitudes as a function of colour, coloured as a function of $\log g$. We can see a turn-off $J-K$ colour of approximately 0.3~mag, equivalent to a turn-off effective temperature of 6400~K. The turn-off colour of a Yonsei-Yale isochrone \citep{YY} with an age of 10 Gyr, a metallicity of $-1.5$~dex, and an \alfe\ of     +0.4~dex was a good representation for our sample. As one can see in Fig.~\ref{fig:HR}, the number of dwarf stars at $J$ $<$ 14~mag is very small. Most of the stars at these magnitudes are redder than the turn-off and have low surface gravities, indicating that these stars are most probably metal-poor halo giants at greater distances. The contamination of our sample by metal-poor thick-disk turn-off stars is very small at these metallicities, as seen by the very few stars which are brighter than the halo turn-off stars. It is worth commenting that the recent study of \cite{2014arXiv1408.3165B} obtained a contribution of the metal-weak thick disk at these metallicities of 25\%. Although Fig.~\ref{fig:HR} suggests that the contamination of thick-disk stars is small in our sample, a full kinematic analysis would be needed to determine the contamination percentage. This would help make consistent comparisons with previous studies of this issue, such as those of \cite{2014arXiv1408.3165B} and \citep{Kordo2013}.

\section{Results}

\begin{figure}[t]
\includegraphics[width=0.99\linewidth]{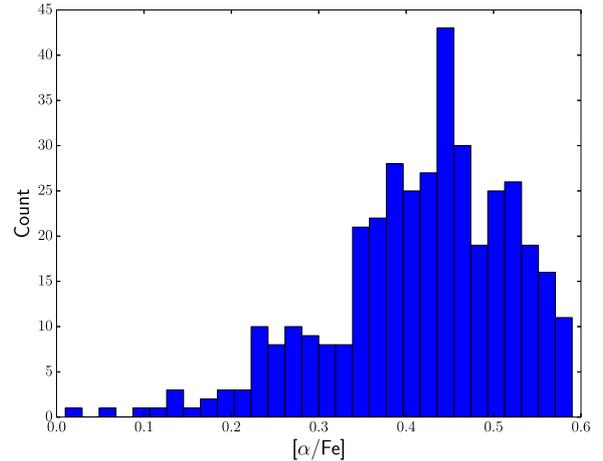}
\caption{ \tiny The distribution of \alfe\ of the metal-poor sample from GESviDR2. An $\alpha_{\mathrm{poor}}$ and $\alpha_{\mathrm{rich}}$ population can be seen. }
\label{fig:hist}
\end{figure}

A distribution of the $\alpha-$abundances of this sample is visualised in  Fig.~\ref{fig:hist}. Our sample is not represented by a single Gaussian distribution, since a second population in \alfe\ below about +0.3~dex is present. This agrees with \cite{N10}, so we chose  a \alfe\ value of +0.3~dex as the boundary between the $\alpha_\mathrm{poor}$ and $\alpha_\mathrm{rich}$ populations. We tested boundary variations of $\pm$ 0.05~dex around +0.3~dex, which did not change the conclusions of our work.

To account for the errors in the parameters, we  propagated them using Monte Carlo simulations. The \alfe\ value of each star was modelled as a Gaussian distribution with a mean value and standard deviation given by the \alfe\ value and its error. Then, by randomly selecting an \alfe\ for each star and counting the number of stars about the \alfe$ = +0.30$  threshold, a value of the \ratio ratio was found. This method was repeated 10,000 times and the mean ratio found along with its standard deviation.

\subsection{The \ratio ratio}

\begin{figure}[t]
\begin{center}
\begin{tabular}{@{}cc@{}}
	
    \includegraphics[width=.25\textwidth]{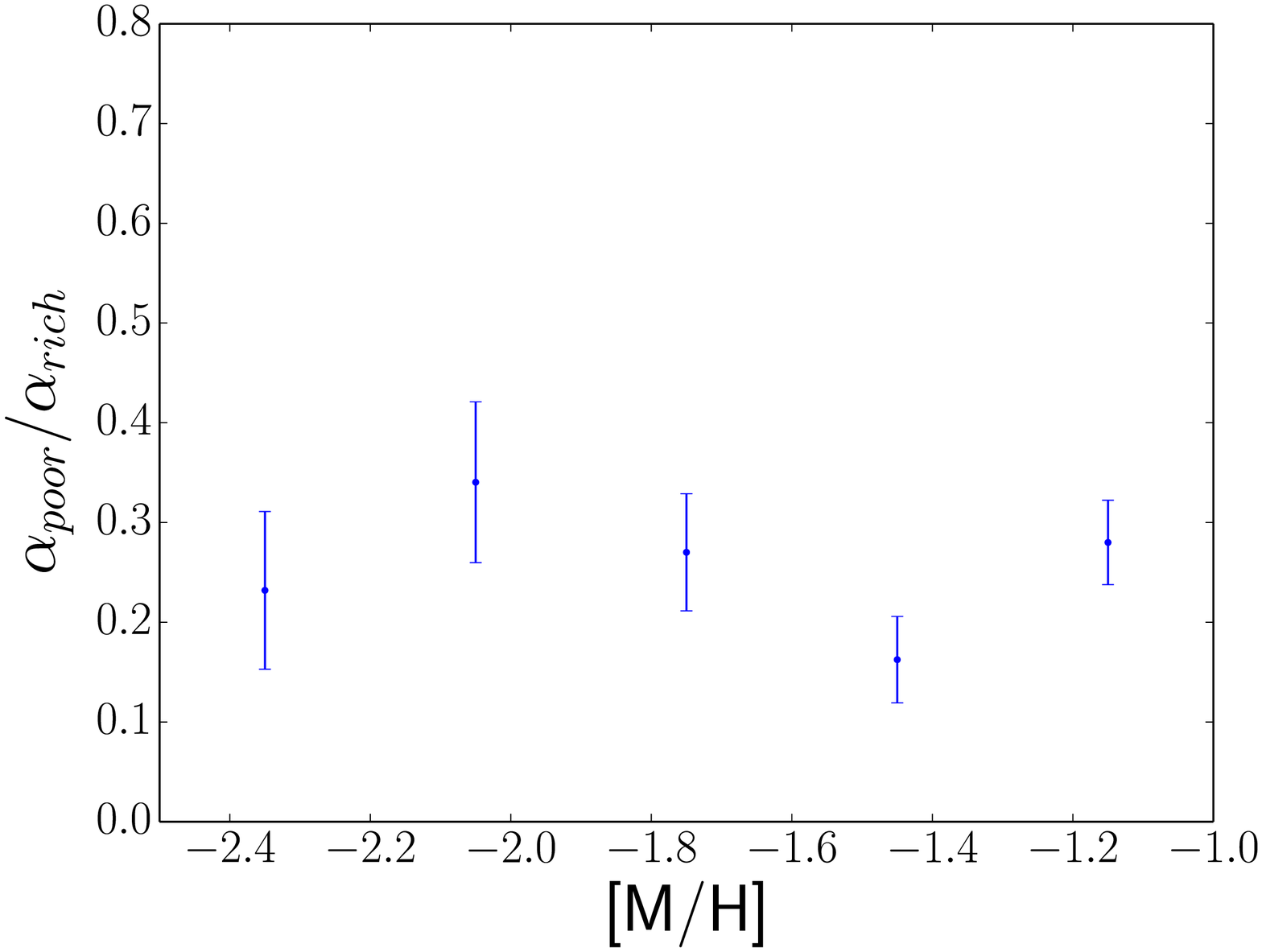} &
    
    \includegraphics[width=.25\textwidth]{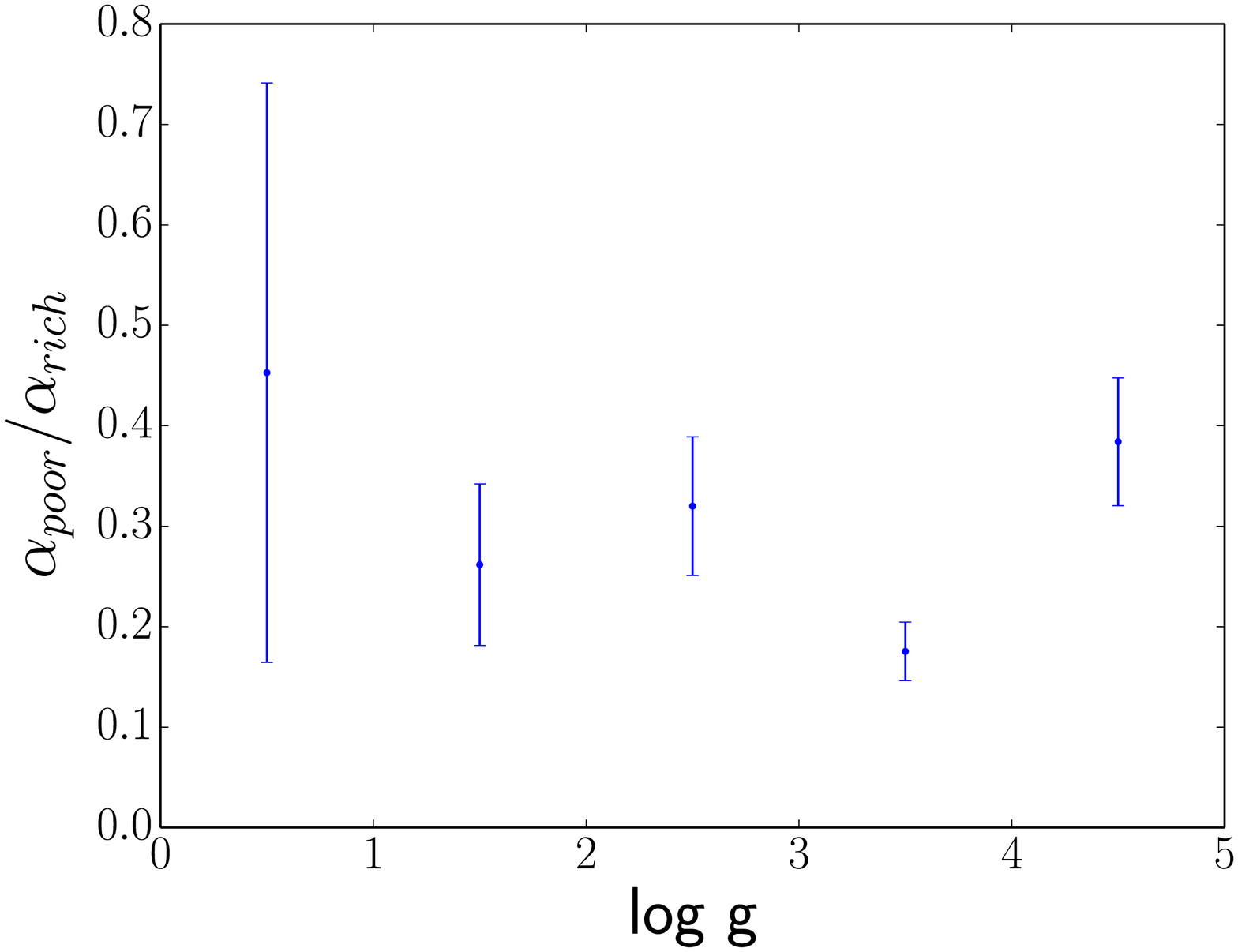} \\
   
    \includegraphics[width=.25\textwidth]{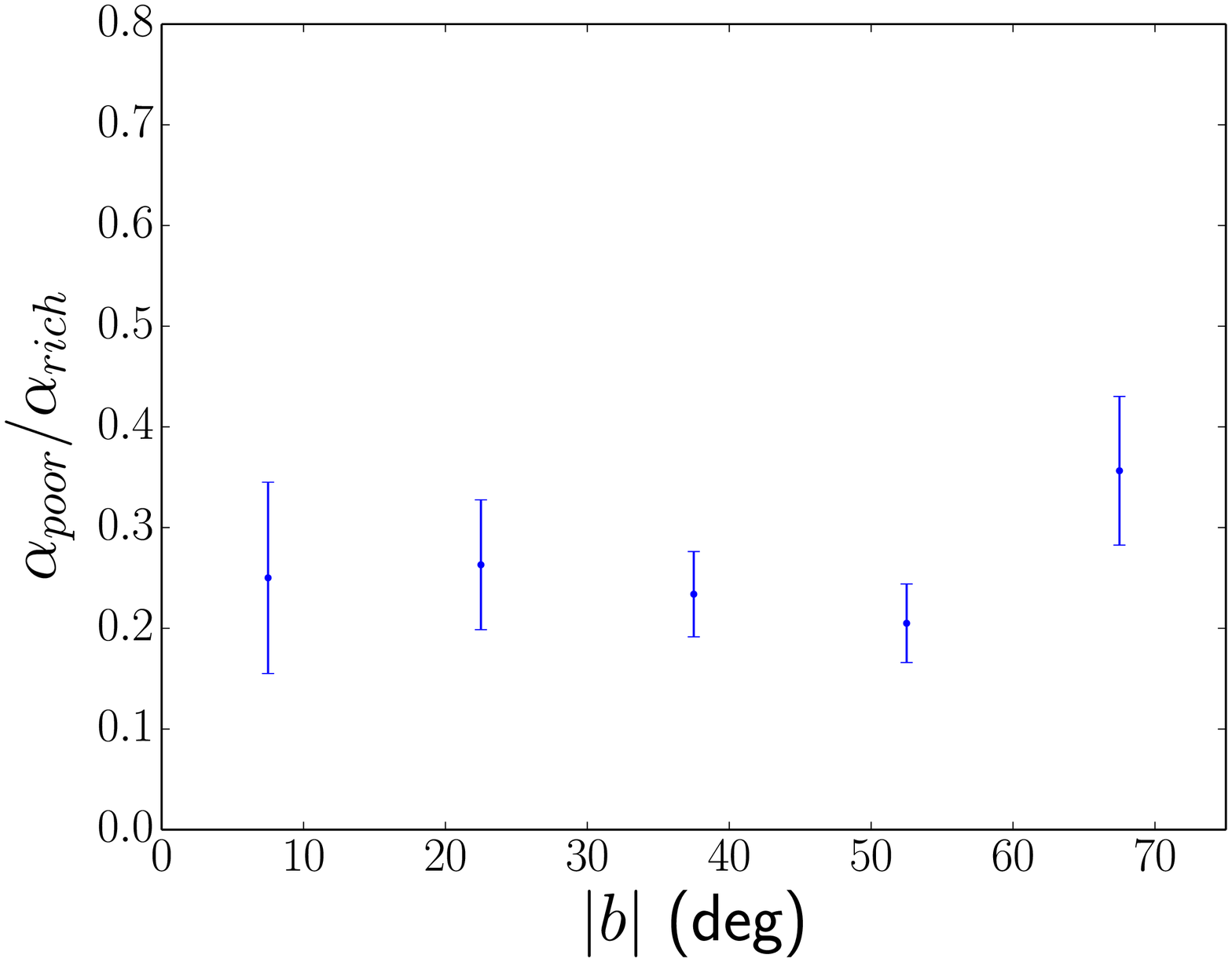} &
  
    \includegraphics[width=.25\textwidth]{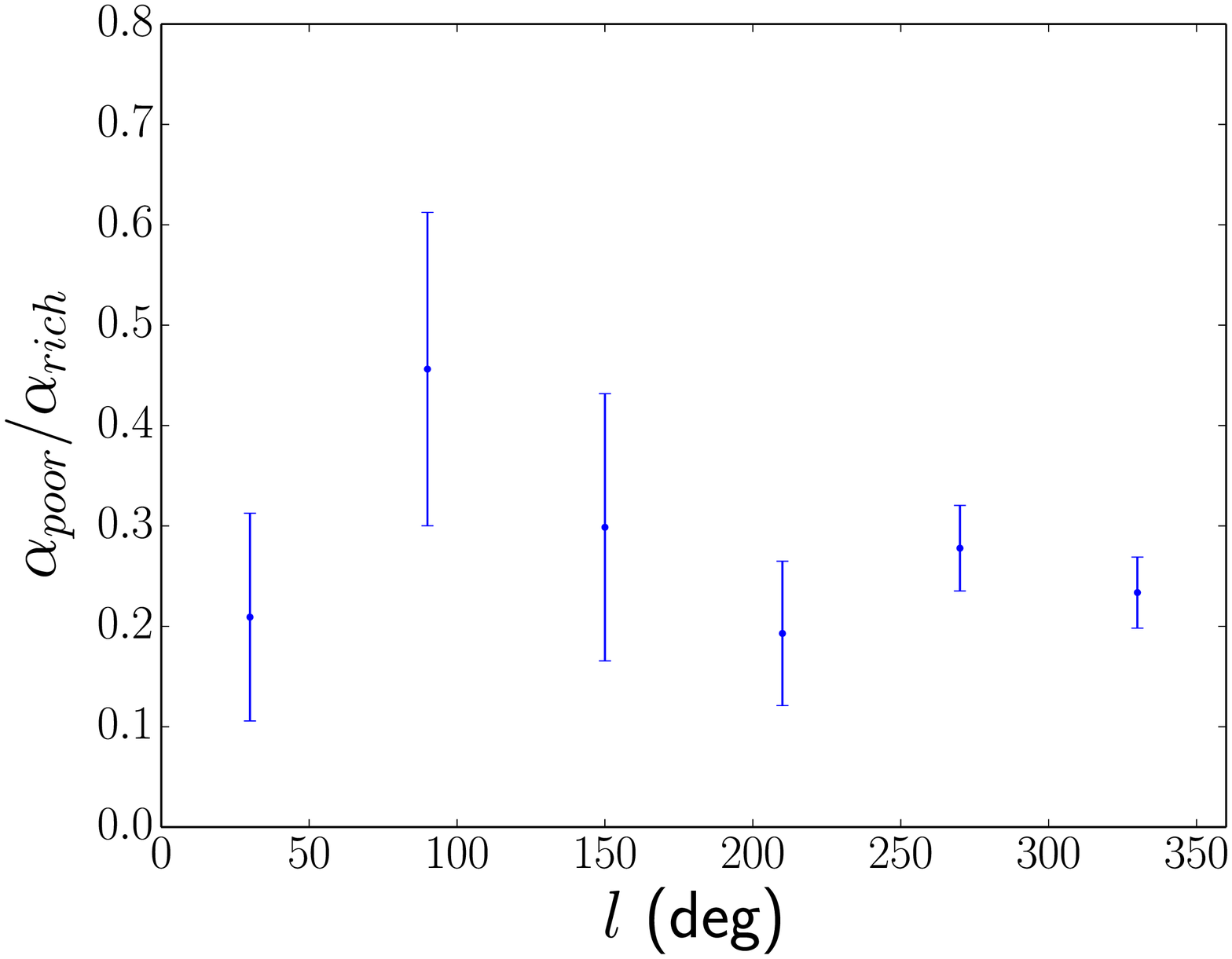}  
\end{tabular}
\end{center}
\caption{\tiny Clockwise, from top left: \ratio ratio as a function of \mh with a bin size of 0.3~dex; \ratio ratio as a function of log $g$ with a bin size of 1.0~dex; \ratio ratio as a function of Galactic longitude with a bin size of 60$\degree$; \ratio ratio as a function of absolute Galactic latitude with a bin size of 15$\degree$.}
\label{fig:ratio}
\end{figure}

The \ratio ratio was determined as a function of metallicity, log $g$, Galactic longitude, Galactic latitude, height above the Galactic plane, and Galactocentric radius as can be seen in Fig.~\ref{fig:ratio} and the top panels of Fig.~\ref{fig:double}.  We  investigated variations in the \ratio ratio with Galactic coordinates in case of overdensities of $\alpha_\mathrm{poor}$ stars. We used $\log g$ to look for biases in the stellar parameters. Metallicity allowed us to determine how the \ratio ratio changes with the metallicity of the formation environment of the stars themselves. 

In Fig.~\ref{fig:ratio} the \ratio values are illustrated. The bin sizes for these parameters were chosen to be at least twice the mean error in the parameter concerned \citep{jofre}.  The top left panel in the figure shows the \ratio\ ratio as a function of metallicity,  showing a mean value of  $0.251 \pm 0.18$. The top right panel shows the \ratio\ ratio, this time with the sample binned in $\log g$, where the mean value is $0.32 \pm 0.10$. The bottom panels show the distribution of \ratio\ as a function of Galactic latitude and longitude, with mean values of $0.26 \pm 0.05$ and $0.28 \pm 0.09$, respectively. Although the \ratio value shows in some cases a slightly different value with respect to the mean, it is not possible to ensure that these have a physical meaning given the small number of stars in some of these subsamples. The bin at low gravities, for example, has three stars only, and the bin at $l = 100$\degree\  has 26 stars. Furthermore, these subsamples do not correlate with each other. For example, the bin at $l \sim 100$\degree\ has all possible values of metallicities, $\log g$, $b$, and radial velocities.  The error bars obtained for the \ratio\ value for each subsample, together with the lack of a systematic trend  or clump for this value, allow us to compute the ratio for the whole dataset obtaining a value of \ratio ~$=0.27 \pm 0.08$.  

\subsection{Stellar distances}
Using the method outlined in \cite{kordo1aa, Kordo2013aa}, we derived stellar distances. We  used those stars whose 2MASS  $J$ magnitudes \citep{2mass} and the associated errors  were known, whose combination of parameters could be utilised by the  method of \cite{kordo1aa,Kordo2013}. 
This left us with  357 stars with derived absolute magnitudes. The distance modulus, $\mu$ = \emph{m$_J$ - M$_J$}, was calculated for each star and the uncertainties in the calculated absolute and apparent magnitudes were propagated to the distance modulus.
The error in the distance was calculated using the standard error propagation formula.
We excluded stars from our sample whose fractional distance error was greater than 0.6 leaving us with a subsample of 291 stars. 

Using the distances and the Galactic coordinates  we calculated Galactocentric radius R and vertical height  above the  plane Z. In the same way as in Fig.~\ref{fig:ratio}, we calculated the \ratio ratio as a function of these parameters for the 291 stars with 1~kpc-sized bins except the last two which have 2~kpc-sized bins. We then computed the mean \alfe\ as a function of Galactocentric radius and absolute plane height. These bin sizes were chosen to ensure there was a sufficient  number of stars to determine the error bars.

\begin{figure}[t]
\hspace{0.3cm}
\includegraphics[width=0.99\linewidth]{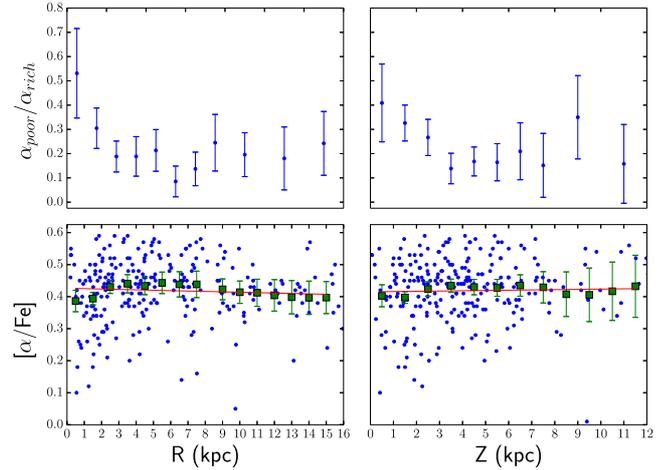}
\caption{\tiny Top Panels: The \ratio ratio as a function of absolute Galactocentric radius (left) and Galactic plane height (right).
Bottom Panels: The 291 stars with fractional distance errors less than 0.6 are shown in blue. The mean \alfe\ value for each 1~kpc R and Z bin and its error is shown in green in the left and right panels, respectively. A linear fit for these mean \alfe\ values is shown in red.}
\label{fig:double}
\end{figure}

{A slight trend can be seen in the top right panel of Fig.~\ref{fig:double}, with the \ratio ratio decreasing slowly as a function of Z for stars for stars within 8 kpc.  Closer to the plane ($|\mathrm{Z}| < 2$ kpc) the \ratio\ has an absolute value significantly larger than the value farther away from the plane. From $|\mathrm{Z}| > 2$ kpc the \ratio\ value is slightly below 0.28 as found generally in Fig.~\ref{fig:ratio}.  However, the error bars do not allow us to fully separate the absolute value of the \ratio\ ratio i.e. inside and outside of $|\mathrm{Z}| = 2$ kpc.  Furthermore, at $|\mathrm{Z}| > 2$ kpc the trend disappears with the ratio appearing to be constant, within the errors. However, this trend is not seen as a function of R in the top left panel of Fig.~\ref{fig:double}, where  within the errors the \ratio ratio is more or less constant except for the first bin (which is very uncertain as it contains only 18 stars).  To conclude that the step in Z is real, more stars with more accurate distances are needed to decrease the bin size of distances and the error bars. In the bottom panel of Fig.~\ref{fig:double} we can see that within the errors of our mean \alfe\ values, the \alfe\  value is essentially constant for stars at different distances. This gives credence to the idea that the stars of the inner halo formed from large proto-galactic clouds with some contamination by accreted material of smaller subunits. We note that with further GES data releases it would be possible to extend the reach of this \ratio determination to the stars above 12~kpc, where the \ratio could differ from the value found for the inner halo.
 
\section{Discussion}
Early origin scenarios for the halo were proposed by \cite{Eggen} and \cite{Searle}, each predicting different properties of the Milky Way and its halo. However, current observations and $\Lambda$CDM (Lambda Cold Dark Matter) models suggest that neither scenario is entirely correct, with the rapid collapse of \cite{Eggen} being insufficient to explain the substructure observed in the outer reaches of our Galaxy by \cite{Bell} and the accretions of \cite{Searle} failing to explain the small metallicity gradients observed by \cite{Dauphole}. The observations seemed to imply that a combination of the two extremes was a better way to explain the formation process. 

The simulations of \cite{Zolotov, Zolotov2}, \cite{Font}, and \cite{2013MNRAS.432.3391T}, for example,  produce galaxies from singular gas collapses combined with accretion aiming at understanding the importance of those processes in the formation history of the halo.  They track this history using the so-called \emph{in situ} and accreted stars. We interpret the \emph{in situ} stars as being $\alpha_\mathrm{rich}$ and the accreted stars as being $\alpha_\mathrm{poor}$ \citep{N10}.
 The value of the ratio of accreted to \emph{in situ} stars is computed in simulations using different input parameters. These parameters include the mass of the collapsing gas, the mass of accreted gas, the mass of the dark matter, the effects of supernova feedback, and the merger history. The ratios of the accreted stars to the \emph{in situ} stars is predicted to be 0.5 \citep{Font}, in the range of 0.4 - 4.0 \citep{Zolotov}, or at least 0.15 - 0.7 \citep{2013MNRAS.432.3391T}. In general, there is a combination of \emph{in situ} and accreted stars at short distances, but at large distances there is an agreement that the halo stars are mostly accreted stars. Our results come from observations at distances smaller than those modelled by most of the simulations. These small distances are also the most uncertain regions in the simulations. Our \ratio ratio of $0.28 \pm 0.08$ should be used to constrain these models. Supernova feedback and the mass of the accreted sub galactic systems, for example, were highlighted as important factors in determining the accreted to \emph{in situ} ratio \citep{Zolotov, 2013MNRAS.432.3391T}.

Other observational campaigns have attempted to quantify the ratio between accreted and \emph{in situ} populations in the Galactic halo.  \cite{Sheffield}, for instance, used a large sample of  M-giant stars covering a large region of the sky. They find and equal number of {\it in situ} to accreted stars in the halo, in better agreement with some of the above simulations than our 0.28 $\pm$ 0.08 result. It is important to note that some of their accreted stars are $\alpha_\mathrm{rich}$ \citep[see Fig. 7 in][]{Sheffield}.  In our case some of the accreted metal-poor stars ([Fe/H] $< -1.5$) could be $\alpha_\mathrm{rich}$  from e.g. Sagittarius \citep{boer}, but if this portion were significant, then we would expect a trend as a function of metallicity in Fig.~\ref{fig:ratio}.  

\cite{Unavane}, on the other hand, using main-sequence turn-off stars found a ratio of 0.1 between accreted and \emph{in situ} stars in the halo. Their conclusions are drawn from the assumption that {\it in situ} stars are coeval presenting a well-defined turn-off. They found that 10\% of their sample was hotter than the turn-off and interpreted that the younger population was the accreted one. It would be interesting to compare future [$\alpha$/Fe] determinations of the \cite{Unavane} sample with our own study. 

Using our distance measurements we can see that these results are focused on the inner halo. We also draw attention to the fact that the mean \alfe\ value does not appear to change dramatically for Z$<$ 12 or R $<$ 16~kpc. The \ratio ratio step at Z = 2 kpc  seen in Fig. \ref{fig:double} needs to be confirmed upon the release of future GES data. For now, we argue that within the errors, we could not find any solid evidence against the claim that a uniform \ratio ratio exists within the inner halo, implying that approximately 85\% of these halo stars formed very rapidly in a high star formation rate environment, such as a high density, uniformly collapsing gas cloud.

\section{Conclusions}
We conclude that the \ratio ratio has a value of $0.28 \pm 0.08$ for the inner halo and is independent of metallicity, log $g$, Galactic longitude and latitude, plane height, and distance from the Galactic centre.  This ratio persists spatially in the solar neighbourhood and there is no clumping of stars as a function of their \alfe\ value. Our result does not depend on $\log g$, thus the \ratio ratio is unbiased between dwarfs and giants. The high $\alpha$ ratio of stars forming at a wide range of [Fe/H] suggests formation of the dominant inner halo population in a high star formation rate environment, with the minority low-$\alpha$ population consistent with formation in a much lower star formation rate (series of) environments.

The second data release of GES contains observations from the first 18 months of the five-year survey and is already showing its potential to quantify the parameters that allow us to deepen our understanding of the Milky Way.
With much more accurate parallaxes and proper motions from the \emph{Gaia} satellite and the remaining data releases of GES,  this work could be extended to the outer halo. With these data we could gain much insight into the importance of mergers with components characterised by low \alfe values in the formation history of the Milky Way.
\begin{acknowledgements}

 Based on data products from observations made with ESO Telescopes at the La Silla Paranal Observatory under programme ID 188.B-3002.
This work was partly supported by the European Union FP7 programme through ERC grant number 320360 and by the Leverhulme Trust through grant RPG-2012-541.
We acknowledge the support from INAF and Ministero dell' Istruzione, dell' Universit\`a' e della Ricerca (MIUR) in the form of the grant Premiale VLT 2012  and  grant 2010LY5N2T.
The results presented here benefit from discussions held during the Gaia-ESO workshops and conferences supported by the ESF (European Science Foundation) through the GREAT Research Network Programme.

KH acknowledges the British Marshall Scholarship program and the King's College Cambridge Studentship. TB is funded by grant no. 621-2009-3911 from the Swedish Research Council. AJK and UH acknowledge support by the Swedish National Space Board.
\end{acknowledgements}
\bibliographystyle{aa} 
\bibliography{AAref1}
\end{document}